# Three Pathways of Student-AI Interaction: Constraint-First Design for Higher-Order Thinking

**Fatima T. Zahra**[1]

AIRELab, Department of Educational Leadership and Policy Studies, University of Tennessee, Knoxville[1]

**Abstract**

How students interact with artificial intelligence (AI) systems in educational settings may determine whether that interaction supports or displaces critical thinking. This paper introduces two contributions. The first is the Three Paths of Student-AI Interaction, a typological framework identifying three qualitatively distinct modes of student-AI engagement: Passive Review, Direct Question, and Strategic Dialogue. The second is the Next Level Teaching Blueprint (NLTB), a three-stage instructional design system intended to make Strategic Dialogue more likely. Qualitative content analysis of 50 randomly sampled student-AI interaction messages from an undergraduate research methods course was used to examine the typology. Two human coders achieved 68% path-level agreement ($\kappa = .48$), with 80% agreement on Strategic Dialogue identification specifically. GPT-5, used as a third coder, produced a similar overall distribution and introduced a coding category absent from the human scheme. Path 1 (Passive Review) accounted for 46% of exchanges in the primary researcher's classifications, Path 2 (Direct Question) for 18%, and Path 3 (Strategic Dialogue) for 36%. A second, descriptively examined dataset contained predominantly Strategic Dialogue content, offering a preliminary indication that instructional framing may influence which path students take. Together, the Three Paths framework and the NLTB contribute a language for describing student-AI interaction and a design approach for supporting higher-order engagement.

## Introduction

Higher education's response to generative AI has been, with few exceptions, reactive and binary: permit or prohibit, embrace or resist. This framing has generated considerable institutional energy and remarkably little pedagogical insight. The disconnect between critical thinking aspirations and pedagogical practice in higher education is not new; it predates AI and persists across national contexts (Ennis, 1985; Zahra, 2012). What generative AI has done is make the consequences of that disconnect acute and visible. Policies address what students do with AI, whether they are allowed to use it and in what contexts (Schiff, 2021), while largely deferring the more consequential question: what happens to students' thinking when they interact with AI, and how does the nature of that interaction determine the answer?

The risk that AI integration compounds the critical thinking gap in place of closing it is evident in literature. Overreliance on generative AI displaces analysis and weakens reasoning capacities (Guo & Lee, 2023; Schiff, 2021).

This paper proposes and defends a specific answer. Not all student-AI interaction is equivalent. When a student submits a prompt and transcribes the response, the cognitive and epistemic experience is fundamentally different from what happens when a student uses AI to pressure-test her own reasoning. Both students used AI. Only one did something that looks like learning.

The central theoretical contribution of this paper is a typology I call the Three Paths of Student-AI Interaction, first presented at a major AI conference in 2025. The Three Paths framework identifies three qualitatively distinct modes of student-AI engagement—Passive Review, Direct Question, and Strategic Dialogue—each associated with a qualitatively different mode of engagement.

## Theoretical Foundations

### *Cognitive Offloading and the Stakes of AI Interaction*

A powerful cognitive tool entering a learning environment does not automatically produce learning. The educational technology literature has long understood that the same tool can scaffold deep thinking or displace it entirely depending on how it is used (Jonassen, 1994; Salomon et al., 1991). Generative AI raises those stakes considerably.

Distributed cognition theory helps explain why. When cognitive work is routinely offloaded to an external system, learners have fewer opportunities to practice and therefore may not develop these skills (Hutchins, 1995). Empirical work documents this risk in AI-specific contexts. Bastani et al. (2024) found that students who used AI assistance performed significantly worse once the tool was removed, suggesting dependency in place of learning.

Fan et al. (2024) described "metacognitive laziness," in which students defer judgment about AI output quality and bypass evaluative reasoning. Doshi and Hauser (2024) found that AI-assisted ideation produced more output but reduced the diversity of ideas, as users gravitated toward thematically overlapping suggestions. Kosmyna et al. (2025) provided EEG evidence that the cognitive costs of LLM reliance may accumulate over time. Across three sessions, LLM users showed weaker neural connectivity, lower self-reported ownership of their writing, and poorer recall of what they had written. When reassigned to write without tools in a fourth session, they showed reduced neural engagement compared with students who had written independently throughout. Across contexts, AI use frequency has been negatively associated with critical thinking performance (Gerlich, 2025), repeated AI assistance with reduced epistemic agency (Darvishi et al., 2024; Liu et al., 2026; Tian & Zhang, 2025), and over-reliance on AI dialogue systems with weaker critical and analytical reasoning (Zhai et al., 2024). If the tool does the thinking, the student does not. Messeri and Crockett (2024) describe a related "illusion of understanding," in which users develop confidence in their comprehension of material that an AI system has processed for them.

The question facing higher education is under what conditions engagement with AI supports, and does not displace, students' thinking. Vygotsky's zone of proximal development (ZPD), the space between what a learner can accomplish independently and what she can accomplish with expert guidance, provides a theoretical frame for productive AI use (Vygotsky, 1978). When positioned well, AI can serve as this kind of scaffold, extending what learners can do without doing the reasoning for them. Vendrell and Johnston (2026) make a similar argument, showing that what LLMs do for learning depends on the pedagogical conditions under which they are deployed.

### ***Higher-Order Thinking as the Non-Negotiable Outcome***

Higher-order thinking encompasses the cognitive processes at the upper registers of Bloom's revised taxonomy, analyzing, evaluating, and creating (Anderson & Krathwohl, 2001), together with the metacognitive reflection through which learners monitor and regulate their own reasoning (Flavell, 1979). Critical thinking, defined by Ennis (1985) as reasonable reflective thinking about what to believe or do, is the most extensively studied component of higher-order thinking but not its entirety. What generative AI does, absent deliberate pedagogical design, is exert structural pressure in the opposite direction. The tool is capable of performing cognitive work at every level, including the higher ones, and higher education has not previously faced a tool capable of doing so much of this cognitive work for a learner.

Lee et al. (2025) found that higher AI confidence was associated with less critical thinking engagement while higher self-confidence was associated with more, meaning the students who trust AI most are least likely to interrogate it. In AI-integrated pedagogy, critical thinking is a primary competency at risk and one around which pedagogical design should be organized.

### *AI Literacy and Critical Thinking as Co-Primary Goals*

AI literacy, defined as the ability to critically understand, evaluate, and responsibly deploy AI systems (Long & Magerko, 2020), is theoretically distinct from general critical thinking but structurally intertwined with it in ways that matter for instructional design. High AI literacy means understanding more than how to use the tool. A learner needs to understand where its answers come from, recognize the blind spots in its training data, and know when to trust her own disciplinary judgment over a confident AI response. As a result, AI literacy requires more than exposure to information about AI, and it develops through critical and repeated use in contexts where the tool's limitations become consequential.

The Three Paths typology maps directly onto a developmental trajectory of AI literacy. Path 1 students treat AI as authoritative (low literacy); Path 2 students use it instrumentally with some evaluative judgment (emerging literacy); Path 3 students interrogate its reasoning and maintain epistemic agency throughout (high literacy). AI literacy and critical thinking are co-primary, mutually reinforcing goals: a pedagogy that develops one without the other will be insufficient.

### *Critical Pedagogy and Epistemic Agency*

Freire's critique of the 'banking model' of education finds a new application here (Freire, 1970). Generative AI, in its most common mode of use, can replicate this  model with algorithmic efficiency where the student submits a prompt and transcribes the output, substituting one depositor for another. The NLTB is designed so that no stage permits passive receipt of AI output. The student is positioned as the epistemic agent throughout (Hofer & Pintrich, 1997; Kuhn, 1999), with AI serving as a resource instead of an authority.

## The Three Paths of Student-AI Interaction

I introduce the Three Paths of Student-AI Interaction as an original typological framework for understanding how students engage generative AI in educational contexts. Developed through systematic theoretical analysis of the cognitive science, educational technology, and AI literacy literatures, the typology was subsequently examined through the pilot study described in Section 4. To the best of my knowledge, it is the first framework to identify these three patterns of student-AI engagement and link them systematically to differences in cognitive and epistemic agency.

The typology distinguishes three paths—Passive Review, Direct Question, and Strategic Dialogue—the degree of student cognitive agency and the nature of the AI interaction. Three pairs of composite student personas ground each path concretely: Aisha and Malik (Path 1), Kofi and Amara (Path 2), and Carmen and Jamal (Path 3). These are theoretically representative composites derived from observed pilot patterns, not individual participants. Figure 1 presents the typology visually; Table 1 provides a systematic comparison across all dimensions.

**Table 1**

*The Three Paths of Student-AI Interaction (Author, 2025)*

| Path | Name | Student Behavior | Mode of AI Engagement | Opportunity for Higher- |
|---|---|---|---|---|

| | | | | Order Thinking |
|---|---|---|---|---|
| **Path 1** | Passive Review | Student accepts AI output with minimal scrutiny; defers to AI as authority. | Unidirectional. Student receives; does not interrogate. | Limited |
| **Path 2** | Direct Question | Student queries AI to fill discrete knowledge gaps; evaluates responses at surface level. | Transactional. Query and response with limited follow-through. | Moderate |
| **Path 3** | Strategic Dialogue | Student engages AI as a thinking partner; probes AI reasoning, challenges outputs, compares AI thinking to her own. | Dialogic. Recursive, critical, and metacognitive. | Substantial |

A diagram showing three paths from Student to AI: Path 1 Passive Review offers limited opportunity for higher-order thinking, Path 2 Direct Question offers moderate opportunity, and Path 3 Strategic Dialogue offers substantial opportunity.

**Figure 1**

*The Three Paths of Student-AI Interaction (Author, 2025). Composite student personas illustrate each path: Aisha and Malik (Path 1: Passive Review), Kofi and Amara (Path 2: Direct Question), and Carmen and Jamal (Path 3: Strategic Dialogue).*

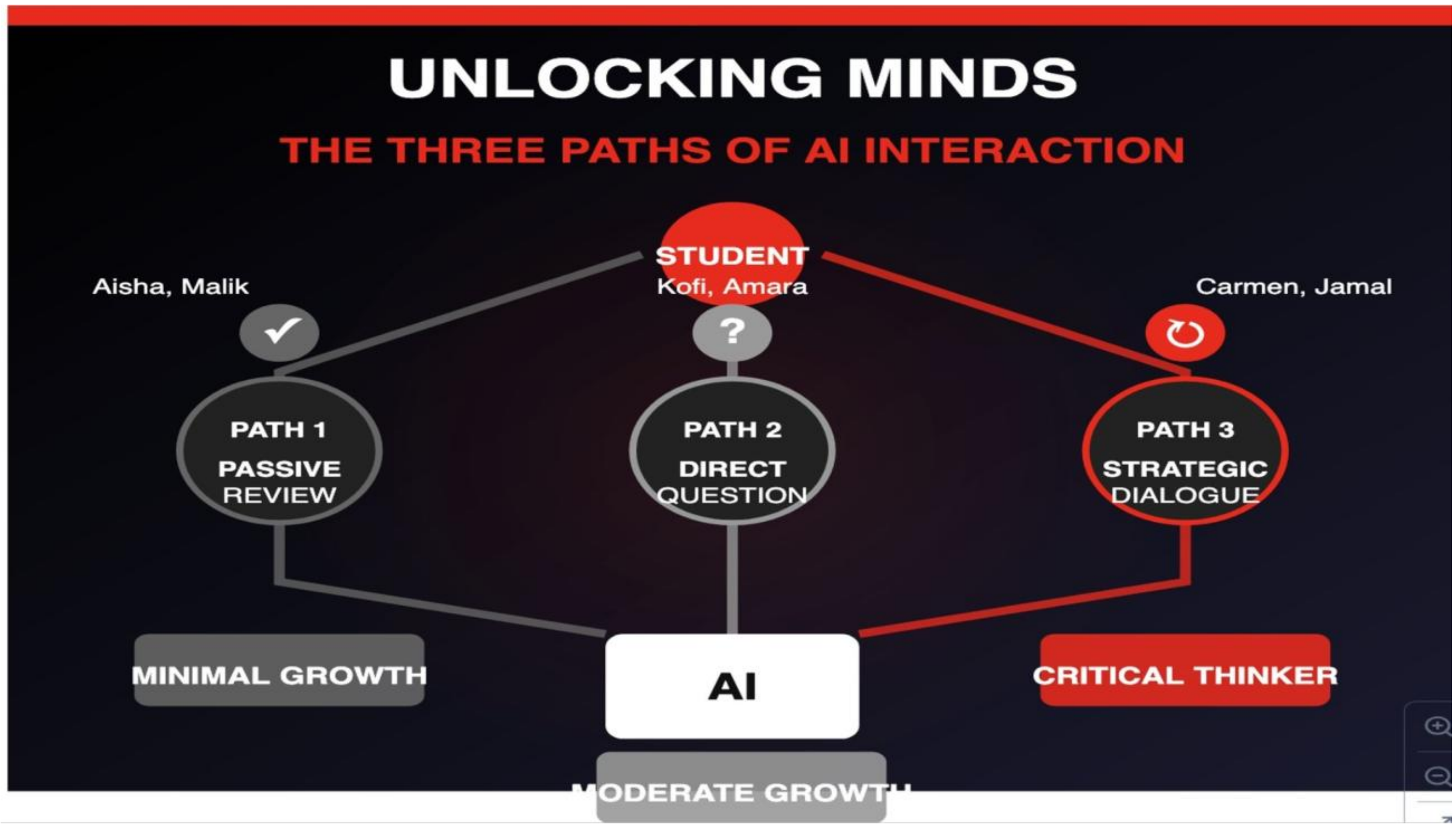

***Path 1: Passive Review***

Aisha and Malik use AI the way most students use it when left to their own devices. They submit a prompt, receive output, and move on. AI is experienced as a knowledgeable source whose answers are presumed correct unless obviously absurd, an oracle with better grammar than Wikipedia and faster response times than a TA. The cognitive investment required is minimal. So is the cognitive development.

Path 1 is more often a literacy deficit than a moral one. Students who do not understand that large language models generate probabilistic text not verified claims, and have no basis for interrogating their outputs. The AI has done the work; the student has received the product. That transaction provides little opportunity to exercise critical thinking.

***Path 2: Direct Question***

Kofi and Amara represent the next level of sophistication and, in our experience, the most common interaction pattern among students in AI-integrated coursework. They know what they do not know, or at least they know what they need. They formulate a targeted question, evaluate the response against their prior understanding, and integrate the answer into their work. This is genuinely more cognitively demanding than Path 1. Question formulation requires metacognitive awareness, knowing what you do not know is no trivial thing.

But Path 2 remains transactional. Students leave with a gap filled, not with the gap examined or the reasoning interrogated. It requires more cognitive engagement than Path 1 but provides limited opportunity for sustained evaluation or revision of one's reasoning.

***Path 3: Strategic Dialogue***

Carmen and Jamal come to AI with a point of view of their own, something Aisha, Malik, Kofi, and Amara do not yet have. They share their own reasoning including their analysis, working hypothesis, or half-formed argument, and ask AI to push back on it. They probe the assumptions in AI's responses. They follow up when something sounds too simple or confident. They ask what perspectives might be missing. They use the exchange not to fill a gap in their knowledge but to stress-test what they already think. The AI in these interactions is not answering the question because it is being utilized to expand the space within which the student has to reason.

Path 3 is cognitively demanding in ways Paths 1 and 2 are not. It requires having a position before the dialogue begins; something to defend, compare, and revise. It also necessitates evaluating what AI produces against one's own reasoning in place of directly absorbing it. The synthesis at the end must show intellectual movement, not repeat what the AI said. These are acts of analysis, evaluation, and synthesis—the upper levels of Bloom's taxonomy. Yang and Ma (2025) similarly distinguish epistemic relationships in human-AI interaction. Yatani et al. (2024) use the term "extraherics" for AI systems that prompt people to examine, instead of replacing, their own reasoning. Path 3 reflects this kind of engagement.

The core claim of the Three Paths typology is that the learning outcomes of AI-integrated education are not primarily determined by whether students use AI, but by which path they are on when they do.

## Methods

### *Design*

This paper reports a qualitative content analysis of student-AI interaction logs from an undergraduate research methods course at an R1 university in the South of U.S.A, conducted to examine the Three Paths typology and characterize how students engage AI in a learning context. The course incorporated AI interaction protocols derived from the developing NLTB framework, giving students structured but open-ended access to a generative AI teaching assistant throughout the semester. The interaction logs provide a naturalistic record against which the typology can be examined.

### *Participants and Setting*

Participants were undergraduate students (n = 14) enrolled in a research methods course at the Disciplinary backgrounds were diverse, spanning education, marketing, forensic anthropology, public health, social work, and related fields, reflecting the interdisciplinary composition typical of an undergraduate research methods course. The course was taught by the author across one full semester. AI interaction was structured according to protocols derived from the Three Paths typology and the developing NLTB framework. All participants provided informed consent under IRB-approved procedures. Research analyses were conducted retrospectively using instructional artifacts generated throughout the course, preserving the natural flow of classroom learning.

### *Data Sources*

The primary dataset consists of 50 student-AI interaction messages randomly extracted by a research team member not involved in the coding process from the full corpus of 580 student-AI interaction messages across seven course assignments produced by 14 students across the semester. Each message represents a complete student turn in a human-AI exchange, the student's prompt to the AI teaching assistant. Random sampling from the full interaction corpus was intended to capture the range of engagement types occurring across the course, not selected examples. A second dataset of 57 messages from 10 students in the same course, drawn from a different assignment context within the semester, provided contextual contrast; it was examined for content patterns but not formally coded.

### *Typology Validation*

Two human researchers, the author (R1) and a trained undergraduate research assistant (R2), independently coded all 50 user messages from the primary interaction dataset. R2 was trained on the codebook through worked examples and practice coding of sample messages. The codebook comprised ten coding labels: six mapping to Path 1 (general_improvement_request, appreciation, assignment_description, course_description, general_context, error_prompt), one to Path 2 (specific_improvement_request), and three to Path 3 (reflection_analysis, applicability_of_method, higher_order_thinking). GPT-5 subsequently introduced two additional labels (calculation_check, citation_check), bringing the full codebook to twelve items; these are described in Finding 3. The codebook is provided as supplementary material. When a message received multiple labels, the first label assigned by each coder determined its path classification.

Each message was assigned a primary label (e.g., reflection_analysis, general_improvement_request, specific_improvement_request), which was then mapped to a path classification. Path-level inter-rater agreement between R1 and R2 was 68% (κ = .48), which Landis and Koch (1977) classify as moderate. Path 2 (Direct Question) was involved in 11 of the 16 discrepancies, consistent with its position as the most ambiguous category in the coding scheme. Coders agreed on Path 3 (Strategic Dialogue) classification in 80% of cases (40/50). Final path classifications were determined by R1, who had designed the course and had contextual knowledge of the instructional setting. As a supplementary check, the majority-vote distribution across all three coders (R1, R2, and GPT-5) was computed separately: P1 = 48%, P2 = 16%, P3 = 36%. This distribution closely mirrors R1's classifications (P1 = 46%, P2 = 18%, P3 = 36%), providing independent corroboration that the reported proportions are not an artifact of a single coder's judgment (Krippendorff, 2018).

As a supplementary triangulation step, GPT-5 was used as a third independent coder. GPT-5 applied the existing codebook but introduced an additional category, calculation_check, for messages in which students submitted completed work and asked specifically about computational accuracy (28% of messages). This category does not exist in the human coding scheme. Human coders distributed those same messages between specific_improvement_request and general_improvement_request depending on the specificity of the query. The scheme divergence reduced direct cross-coder comparability, producing low pairwise kappas between GPT-5 and human coders. GPT-5 identified 28% of exchanges as Path 3, compared with 36% for R1 and 24% for R2. This pattern suggests that Path 3 is detectable across coders, although the scheme divergence limits direct comparison.

### *Analysis*

Interaction log data were analyzed using qualitative content analysis (Krippendorff, 2018), with each message classified against the Three Paths codebook. Path distributions were computed from R1's final classifications, with the three-coder majority-vote distribution used as a supplementary comparison.

## Results

### *Finding 1: The Three Paths Are Empirically Distinguishable*

Analysis of 50 double-coded student-AI interaction messages indicated that exchanges could be classified into three patterns corresponding to the typological paths. Using R1's final classifications, Path 1 (Passive Review) accounted for 46% of coded exchanges, Path 2 (Direct Question) for 18%, and Path 3 (Strategic Dialogue) for 36%. The majority-vote distribution across all three coders (R1, R2, and GPT-5) was very similar, with 48% in Path 1, 16% in Path 2, and 36% in Path 3. GPT-5 also used a calculation_check subcategory, but the overall distribution remained consistent with the typology's core structure.

Paths 1 and 2 together accounted for 64% of student-AI interactions in the primary sample. This is consistent with the typology's prediction that Strategic Dialogue is unlikely to be students' default mode of engagement. A second dataset of 57 messages from 10 students in the same course, drawn from a different assignment structured around AI-prompted reflection tasks, showed a different pattern, with most messages displaying applicative and policy-critical engagement consistent with Path 3. Because this dataset was not systematically coded, the contrast offers only a preliminary indication that instructional framing may influence path engagement. In a separate study of AI-assisted brainstorming in graduate education, Shah et al. (2024) found that only 13% of students critically evaluated and built upon AI recommendations, while 36% accepted aligned suggestions without interrogation and the remainder engaged at varying levels of passivity. Yang and Ma's (2025) classification of epistemic relationships in human-AI interaction also converges conceptually with the distinction between treating AI as a validator and engaging it as a co-thinker. Together, these studies suggest cross-study consistency, although direct comparison requires caution.

***Finding 2: Instructional Framing Shifts Path Distribution***

A second dataset of 57 interaction messages from 10 students in the same course, drawn from a different assignment within the semester, offers a meaningful contrast to the primary sample. Where the randomly sampled dataset captured open-ended AI access with general guidance but no explicit Path 3 protocols, this second assignment structured AI interaction around scaffolded reflection tasks, prompts that required students to engage AI with hypotheses, policy implications, and applicative reasoning over answer-checking. The content-level difference between the two datasets is striking. The primary dataset is dominated by messages seeking validation or correction, 'Are my answers correct?', 'Is there anything inaccurate?', consistent with the 64% Path 1/2 distribution. The second dataset is dominated by messages that probe, extend, and apply: 'How might z-scores help counseling services prioritize which students require immediate support?', 'Do you think income is related to digital literacy?', 'What story do the standardized residuals tell about gender distribution?' These are Path 3 interactions by design. The tasks required this mode of engagement, and students delivered it.

This contrast is offered as evidence that students can produce Path 3 interactions when an assignment requires them to do so. This study is not a controlled comparison; the assignments differed in content and AI prompt structure, making causal claims unwarranted. In the second dataset, the design made Path 1 and Path 2 responses insufficient. This pattern is consistent with the NLTB's design logic.

### ***Finding 3: GPT-5 as a Methodological Partner***

GPT-5 was included to provide another application of the codebook. What proved more interesting was where it departed from the human coders. It introduced calculation_check for messages in which students submitted completed work and asked about computational accuracy. The human codebook had distributed these messages across two improvement-request categories, so the new label made visible a recurring feature that we had not named separately. This was useful analytically even though it reduced direct comparability and produced low pairwise kappas.

The calculation_check category may be especially useful for refining the Path 1/2 boundary. It identifies a recurring type of request that human coders, working from the original categories, had classified differently. Future coding work can test whether retaining this subcategory improves clarity and agreement.

The episode provides an interesting illustration of the paper's broader argument. Used only as a verification tool, GPT-5 agreed imperfectly with the human coders. What we did not expect was for GPT-5 to identify a pattern the human coders had not considered. Working from the same codebook and the same fifty messages, it identified a category, *calculation_check*, that neither human coder had named. We had coded those messages under two separate labels without recognizing the common structure across them. GPT-5 had not been asked to revise the codebook, only to apply it. In a small but interesting way, this reflects the broader argument of the paper. The structure of a task can directly inform what an interaction produces. This was a single coding exercise, however, and should not be taken as evidence that the typology generalizes beyond student–AI interaction.

## The Next Level Teaching Blueprint

The Three Paths typology raised a concrete design question. If Path 3 creates stronger opportunities for higher-order thinking but is not students' usual mode of engagement, what instructional architecture might support it? The Next Level Teaching Blueprint (NLTB), a replicable instructional design system with three sequenced stages, explicit interaction protocols, and a theoretical logic linking each element to Path 3 engagement, is my answer. The NLTB was first implemented in this study. The study's stage design is presented in Table 2.

**Table 2**

*The Next Level Teaching Blueprint: Stage Design and Alignment to Three Paths (Author, 2025)*

| NLTB Stage | Student Activity | Pedagogical Mechanism | Alignment to Three Paths |
|---|---|---|---|
| **Stage 1: Independent Work** | Engage the problem fully without AI. Document reasoning, hypotheses, and uncertainties. | Secures epistemic agency before AI contact. Creates the cognitive anchor for Stage 3 comparison. | Structurally prevents Paths 1 and 2 by requiring students to have their own position before encountering AI output. |
| **Stage 2: Dialogic AI Partnership** | Engage AI dialogically: share reasoning, ask AI to challenge it, probe AI responses, explore alternatives. Answers are not the goal. | Operationalizes Path 3 interaction. Explicit protocols scaffold strategic dialogue as a learnable practice. | Directly trains and rehearses Path 3 behavior. Students cannot execute Stage 2 correctly without engaging in Strategic Dialogue. |
| **Stage 3: Reflective Synthesis** | Return to Stage 1 documentation. Compare original thinking to AI-augmented thinking. Construct an integrated, evaluated position. | Demands metacognitive evaluation (upper-register Bloom) and consolidates intellectual development across the cycle. | Completes the Path 3 loop. Students must adjudicate between their own reasoning and the AI's: evaluate, not summarize or transcribe. |

***Stage 1: Independent Work***

Students must engage the problem fully before any AI interaction is permitted, documenting their reasoning, hypotheses, and uncertainties. Stage 1 exists because students cannot engage in genuine Path 3 dialogue if they have nothing of their own to bring to it. Without Stage 1, Stage 2 collapses into Path 1 or Path 2 by default. Stage 1 documentation also gives instructors visibility into unmediated student thinking, a window of increasing diagnostic value in AI-integrated classrooms.

### *Stage 2: Dialogic AI Partnership*

Stage 2 operationalizes Path 3 as a structured practice. Students engage AI under explicit interaction constraints, the defining design feature of the NLTB. They are not permitted to ask for answers, summaries, or drafts. They must engage dialogically by first sharing their Stage 1 reasoning with AI, then asking AI to challenge it, and finally probing assumptions in its responses and exploring alternatives. In practice, the student steers the reasoning; while AI expands the problem space. Disciplinary implementation is flexible within this structural constraint. In every case, Stage 2 is designed so that Path 1 and Path 2 responses are structurally non-compliant.

### *Stage 3: Reflective Synthesis*

Stage 3 returns students to their Stage 1 documentation and requires a metacognitive act: compare original thinking to AI-augmented thinking, articulate where and why thinking changed, and construct an integrated position. Students produce an evaluated account of intellectual movement not a summary of what the AI said. This work asks them to judge the exchange and revise their position, placing evaluation and synthesis inside the assignment, not leaving them implicit.

## Scaling: From Pilot to Campus-Wide Initiative

Following the pilot study, a campus-wide initiative at the author's institution, funded at $22,100 across two implementation phases, currently involves six faculty across three colleges, with a three-year target of 12 to 15 faculty and approximately 350 students. This adoption across disciplines and instructional contexts suggests the framework is transferable, though formal evaluation of the initiative is beyond the scope of this paper.

## Discussion

The Three Paths typology makes a specific claim that how students engage with AI determines the kind of thinking they do. The content analysis indicates that coders can distinguish the three paths with moderate agreement, including 80% agreement on Path 3. The contrast between the primary sample (64% Path 1/2) and the reflection-structured assignment dataset (predominantly Path 3 content) also suggests that instructional design may be an important variable. Figure 2 presents the trajectory of path distributions across all seven course assignments, showing that Path 3 engagement varied with assignment design and in-class scaffolding. The typology describes interaction modes that can shift when the design changes instead of fixed student types.

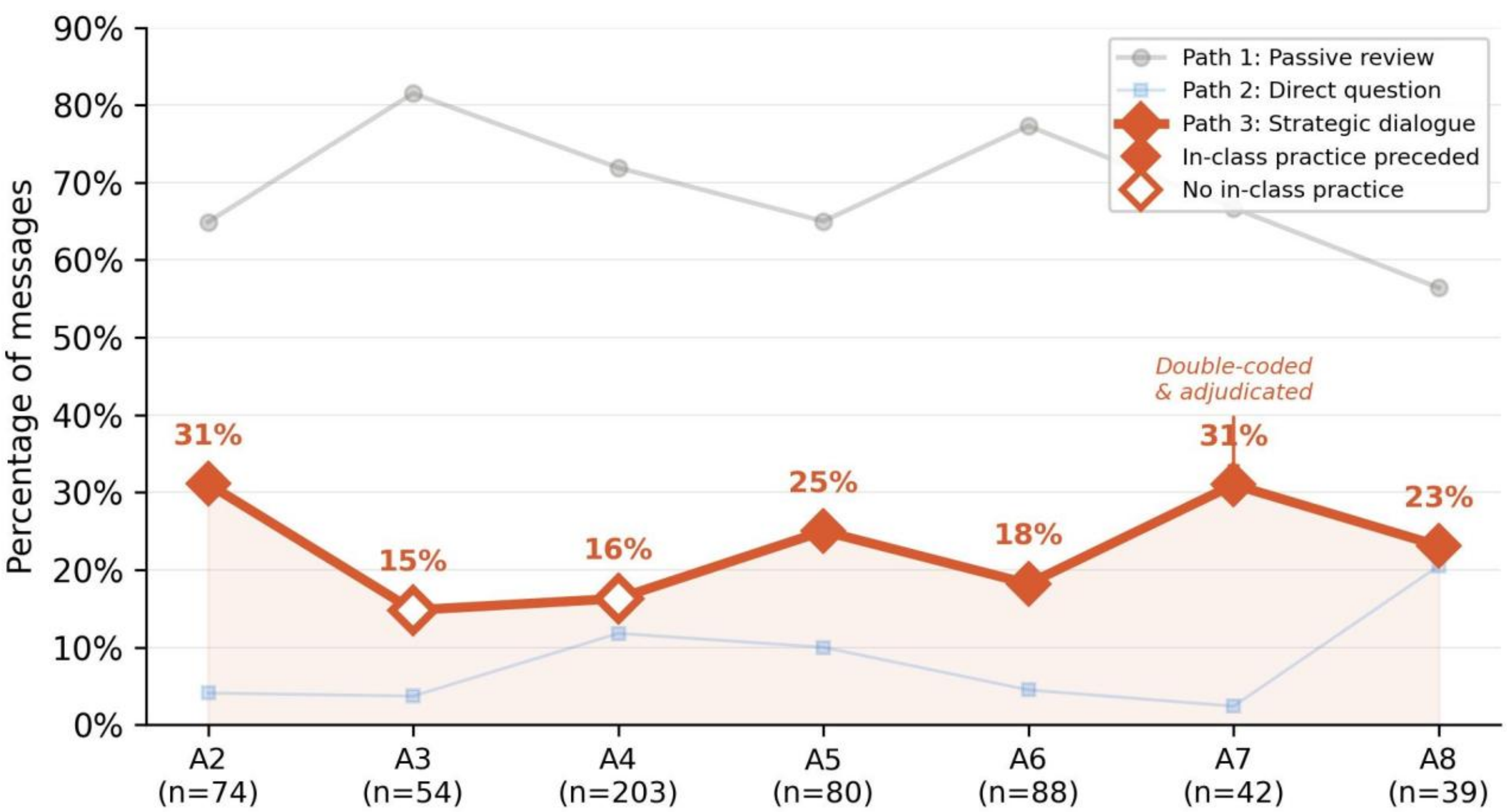

The finding that Paths 1 and 2 accounted for 64% of coded interactions is the most directly actionable result. Without deliberate scaffolding, students default to the interaction modes offering the least cognitive resistance. Path 3 requires effort that students may be unlikely to expend without structural incentives. For faculty, there are practical implications. For example, asking students to 'use AI critically' is not enough as an instructional design because it may as well be a hope. An assignment must require students to arrive with their own reasoning, expose that reasoning to challenge, and account for what changed. Otherwise, a polished answer or a quick verification remains a sufficient response. The NLTB exists because I watched this happen in my own classroom before the framework was conceived.

The GPT-5 finding raises a possibility beyond student-AI interaction. The mode of engagement with AI may influence what the system contributes in other cognitive tasks. This interpretation relies on a single coding exercise and warrants further investigation.

For institutional AI policy, the typological framework reframes the central question. A prohibition policy that prevents Path 1 and 2 harms also forecloses Path 3 benefits. A permissive policy without pedagogical scaffolding produces Path 1 and 2 users at institutional scale. Neither is adequate. The NLTB provides an instructional architecture that makes Path 3 the structural default, by designing the conditions under which students engage AI, not by restricting the tool itself. The policy implication is instructional, not regulatory, since the question is which designs produce thinking, and which do not.

## Limitations and Future Directions

The most pressing question this study cannot answer is whether the cognitive displacement visible in the interaction data translates to measurable differences in what students actually learn. If it does, the cost is not borne by individual students alone as it has institutional consequences. The interaction dataset ($n = 50$ coded messages from 14 students) is small and drawn from a single institution and course context, limiting generalizability. The inter-rater reliability at the path level ($\kappa = .48$) is moderate; the Path 1/2 boundary remains the most contested site of the coding scheme and is a priority for refinement in subsequent work. The second dataset (n = 57, same course, different assignment) was not formally coded and is treated as descriptive context only. This paper focuses on the typological and instructional design contributions; pre-post outcome measurement from the same course context is reported in a separate empirical study and is not presented here.

The immediate priority in this paper is a formal comparison of scaffolded and unscaffolded conditions because it tests the design claim at the center of NLTB. Subsequent work should include larger coding studies across institutions, trials comparing NLTB-designed and conventionally taught courses, refinement of the Path 1/2 boundary, and examination of the typology across first-generation, international, non-Anglophone, and non-US higher education contexts.

## Conclusion

AI use is too coarse a category to guide pedagogy. Two students may use the same system on the same assignment while doing very different cognitive work. The Three Paths framework gives those differences a language, and the NLTB translates that language into a sequence instructors can design around. The pilot data indicate that the three interaction modes can be identified, although the moderate inter-rater agreement and small sample show that the coding scheme needs further refinement. The findings also offer a preliminary reason to examine instructional framing more closely.

This study does not yet carry the weight of large-scale evidence. It does, however, offer a framework for something many instructors already recognize. Students engage with AI in different ways; these differences influence the thinking they do themselves. The framework now needs to be tested across larger samples, disciplines, and settings. In the meantime, treating all

student–AI interaction as equivalent risks overlooking an important distinction. The findings from this pilot suggest that distinction is worth taking seriously. I will continue to test and refine the framework with my research team in our own settings, and hope others will do the same in theirs.